\def\year{2020}\relax
\documentclass[letterpaper]{article} 
\usepackage{aaai20}  
\usepackage{times}  
\usepackage{helvet} 
\usepackage{courier}  
\usepackage[hyphens]{url}  
\usepackage{graphicx} 
\usepackage{graphicx}  
\usepackage{amsmath}
\usepackage{amsthm}
\usepackage{booktabs}
\usepackage{amssymb}
\newcommand{\citet}[1]{\citeauthor{#1} \shortcite{#1}} 
\newcommand{\citep}{\cite}

\title{Moving Metric Detection and Alerting System at eBay}

\author{Zezhong Zhang, Keyu Nie and Ted Tao Yuan \\
eBay Inc., San Jose CA 95125, USA\\
{zezzhang, knie, teyuan}@ebay.com\\
} 
\begin{document}

\maketitle

\begin{abstract}
At eBay, there are thousands of product health metrics for different domain teams to monitor. We built a two-phase alerting system to notify users with actionable alerts based on anomaly detection and alert retrieval. In the first phase, we developed an efficient anomaly detection algorithm, called Moving Metric Detector (MMD), to identify potential alerts among metrics with distribution agnostic criteria. In the second alert retrieval phase, we built additional logic with feedbacks to select valid actionable alerts with point-wise ranking model and business rules. Compared with other trend and seasonality decomposition methods, our decomposer is faster and better to detect anomalies in unsupervised cases. Our two-phase approach dramatically improves alert precision and avoids alert spamming in eBay production.
\end{abstract}

\section{Introduction}
Over the era of big data, large companies have the desire of building automated alerting system to continuously monitor the health of their sites and applications. Currently, many solutions are proposed to serve their needs, including Alibaba \citep{alibaba}, Amazon \citep{amazon}, Anodot \citep{anodot}, Baidu \citep{baidu}, AT\&T \citep{att2012}, Facebook \citep{facebook}, Google \citep{google1}, LinkedIn, Microsoft \citep{ren2019time}, Twitter \citep{hochenbaum2017automatic}, and Yahoo \citep{yahoo}. 

At eBay, product owners monitor various metrics every day. Each metric potentially contains thousands of time series with different dimension aspects. To effectively monitor these millions of time series and alert automatically, we developed a two-phase approach with anomaly detection and alert retrieval to build eBay automated alerting system.

In our practice, there are three challenges for building the automated alerting system:
\begin{itemize}
\item Lack of true alert labels: it is demanding for domain analysts to label past anomalies in each time series for machine to learn,
\item	 Scalability and efficiency: we need to detect millions of time series every day and alert in time,
\item	 Avoid alert spamming: flooded alerts do not help users since not all finding outliers are valid alerts to all users.
\end{itemize}

To tackle the first two challenges, a fast and robust time series anomaly detection algorithm (MMD) is developed. It firstly decomposes a metric time series to "normal" patterns (e.g. trend, seasonality) and a noise/residual part. To gauge the noise level as well as filter out potential anomalies without assuming particular distribution of the noise part, we use Chebyshev’s Inequality to determine the anomaly criteria that controls the overall false positive detection rate (type I error outside “normal” range). At the end of the anomaly detection phase, we obtained a list of time series with anomalies that are potentially used to alert users.

Secondly, to increase precision of the potential alerts and reduce alert spamming, we designed an alert retrieval framework that takes input from the anomaly detection phase, combined with additional rules and policies, to generate the final ranked or prioritized alert list to notify users.

In recent years, many anomaly detectors are proposed to tackle above issues, such as Isolation Forest \citep{liu2008isolation}, Argus \citep{att2012}, OC-SVM \citep{amer2013enhancing}, EGAD \citep{yahoo}, Opprentice \citep{liu2015opprentice}, VAE \citep{an2015variational}, RRCF \citep{amazon}, Donut \citep{alibaba}, SR-CNN \citep{ren2019time} and ATAD \citep{zhang2019cross}. Our time-series anomaly detection focuses more on statistical model as \citet{makridakis2018statistical} pointed out that the performance of ML models is lower to that of statistical methods in time-series area. Besides, many existing anomaly detection methods may cause alert spam since they only focus on anomalies without considering the second phase to retrieve valid alerts. 

In this paper, we first explain our two-phase approach, especially the efficient and robust moving metric detector; then we discuss the evaluation of alert result and production performance; and conclude with some discussion. 

\section{Approach}
\begin{figure}[h]
\centering
\includegraphics[width=0.9\columnwidth]{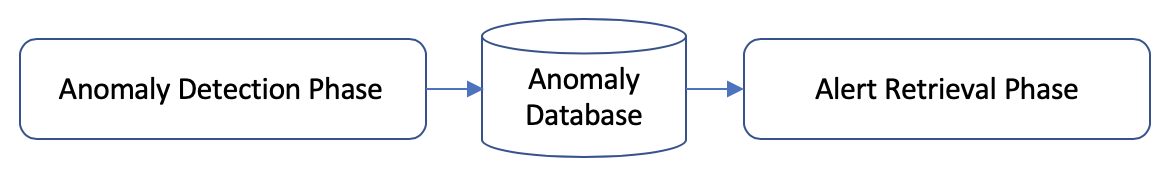} 
\caption{The Two-Phase Alerting System.}
\label{fig1}
\end{figure}

Let’s assume a business metric is measured daily and represented as a time series. Anomaly detection is to determine if the last observation of the metric is significantly different from what it is supposed to be given the past measurements of the time series. In Figure \ref{fig1}, we illustrate the overall alert detection and alert notification pipeline. The anomaly detection phase comprises of time series signal decomposition and statistical test for anomaly. The alert retrieval phase is to filter out valid alert from recently detected anomalies found in the previous phase. 

\subsection{Anomaly Detection Phase}
Our basic assumption is that metric time series data should contain mostly "normal" observations than "abnormal" observations, and we do not require the abnormal data being labeled. In order to identify anomalies, the general scheme is to learn the “normal” patterns in the time series and use it as prediction to test if the last observation is within the “normal” metric value range.

\subsubsection{Time Series Decomposer} 
As key part of anomaly detection phase, we decompose each time series to trend, seasonality and residual parts. There are existing time series decomposition implementations, however, they are either sensitive to outliers because using moving average, e.g. classical decomposition \citep{kendall1983advanced}, or with high computation cost of iterations, e.g. STL \citep{cleveland1990stl}. To extract “normal” patterns from time series containing sparse anomalies in the past but without labels, we developed our moving metric decomposer using median to extract robust trend and seasonality in the time series. 

For a given time series $X$, we first do a rough estimate of trend ($L$) using symmetric moving average with a sliding time window ($w$), which is the number of observations per cycle in the given time-series $X$. To estimate the $w$ for each metric, we applied a one-time signal frequency estimation using ESPRIT method \citep{roy1989esprit} since the frequency of each metric rarely changes for almost all cases in eBay. Using the trend $L$, we can discover seasonality and reduce part of the variance in $X$. 

Let’s define trend-removed time series as $L^{\prime}=X-L$. To extract a robust seasonality ($S$), we use the median of historical periodic values of trend-removed time series $L^{\prime}$, instead of simple averaging which is sensitive to outliers. 
\begin{equation*}
S_{t} = median(L_{t \pm iw}^{\prime} | i \leq \tau, i \in \mathbb{N})
\end{equation*}
\noindent where $w$ is the number of observations per cycle, $i$ is the number of cycles away from observation at $t$ and $\tau$ is the number of cycles near the observation at $t$ that is used to estimate seasonality $S_t$. In our case, we set $\tau$ as length of the input time series $X$ that contains all the observations. 

With extracted seasonality $S$, we then estimate trend once again from the seasonality-removed series as $S^{\prime}=X-S$. We ignore the first estimate of trend $L$ and recompute trend from $S^{\prime}$ as $T$ in the following. Using right alignment to calculate rolling median in a sliding time window ($w$),
\begin{equation*}
T_t^f=median(S_{t-j}^{\prime} | j \leq w, j \in \mathbb{N})
\end{equation*}
\noindent where $j$ is the number of observations previous to observation at $t$. The $f$ indicates it is right alignment.

The right alignment rolling method naturally lags behind current observations, so we estimated a potential bias term using median in the following formula to model the lagging effect. Hence, the extracted trend contains two terms,
\begin{equation*}
T=T^f+median(S^{\prime}-T^f).
\end{equation*}

Compared with STL \citep{cleveland1990stl}, which iteratively estimates trend and seasonality with inner loop and outer loop to moderate outlier impact on trend and seasonality, our method first computes an approximated trend $L$ and extract seasonality $S$, and then extract the final trend $T$ with median as a robust replacement of average. Our method requires less computation with better and robust performance.

\subsubsection{Determine “Normal” Range of a Metric}
With time series decomposer, given time series $X$ is decomposed into trend ($T$), seasonality ($S$) and residual ($R$), which is defined as $R=X-T-S$. We detect anomaly as the most recent observation in a time series whose value differs significantly from the "normal" prediction. To detect anomaly, we have to calculate the “normal” range from the input time series $X$.

In many cases, the noise residual $R$ is assumed to be normally distributed. However, in our approach, we do not assume “normal” residuals distribution, i.e., whether it is Gaussian or not. By using Chebyshev’s Inequality, for any real number $k>0$,
\begin{equation*}
P(|R- \hat{\mu} | \geq k\hat{\sigma}^2 ) \leq  \frac{1}{k^2}
\end{equation*}
\noindent where the $\hat{\mu}$ is the expected value and $\hat{\sigma}$ is the standard deviation of the residual $R$. Using the above inequality formula, we can define false positive criteria to differentiate anomaly values outside of “normal” region. 

The distribution-free “normal” range of $X$ can be calculated with a given expected probability $p$ of seeing an anomaly \citep{amidan2005data},
\begin{equation*}
T+S+\hat{\mu} \pm k\hat{\sigma}
\end{equation*}
\noindent where $k=\frac{1}{\sqrt{p}}$. 

To estimate the $\hat{\mu}$ and $\hat{\sigma}$ for “normal” range, \citet{hampel1974influence} pointed out that mean and standard deviation are sensitive to outliers and \citet{leys2013detecting} suggested using median and median absolute deviation (MAD) as robust replacements, as follow:
\begin{align*}
					 & \hat{\mu} = median(R) \\
					 & \hat{\sigma} = b \times median(|R - median(R)|)
\end{align*}
\noindent where the robust standard deviation $\hat{\sigma}$ is estimated by scaling MAD with a constant $b$ (commonly set as $1.4826$). 

Especially with our decomposition method, we set the expected $\hat{\mu} = 0$ to replace the $\hat{\mu} = median(R)$ due to the extra added bias adjustment by $median(S^{'}-T^f)$. In our model, the k of the “normal” range can be interpreted as the number of standard deviations away from the prediction ($T+S$). With Chebyshev’s Inequality, the “normal” range with $p=0.01$ would be interpreted as at most $1\%$ probability of exceed the range with $k=10$. In this case, we identify anomaly if the value is beyond $10$ standard deviation away from prediction.

\subsection{Alert Retrieval Phase}
Not all anomalies are valid actionable alerts for all users. In eBay practice, we can only send 10 alerts for each user every day, otherwise it will cause alert spamming. In the alert retrieval phase, we developed a ranking model and filtering rules and policies to retrieve valid alerts.

\subsubsection{Ranking of Anomalies}
To retrieve valid alerts and increase precision, we created a point-wise anomaly ranking algorithm with generalized linear ranking model to sort valid alerts from identified anomalies. The ranking score has several components, i.e.,
\begin{itemize}
\item Anomaly deviation severity feature computed using outputs from the previous phase as $f_d=|R-\hat{\mu} |/ \hat{\sigma}$ .
\item User defined 4-level priority (e.g. P1, P2, P3, P4) for each metric as $f_p$ (one-hot encoding).
\item Granularity of time series dimension aspect values as $f_g$, which is the number of dimensions that not rolled up in metric dimension hierarchy. For example, “Country: US” has granularity $f_g=1$ and “Country: US, Device: PC” has granularity $f_g=2$. 
\end{itemize}

With above features, our ranking model is
\begin{equation*}
g(p)=w_d f_d+w_p f_p+w_g f_g
\end{equation*}
where $w_p$, $w_g$ and $w_d$ are the weight vector that can be learned from users feedbacks. The $g(p)$ is the ranking score function considering both anomaly severity and importance.

\subsubsection{Retrieval Logic and Rules}
Some anomalies that we detected are true and valid alerts to users, however, they are related and pointing to the same issue. From user perspective, it is one kind of alert spam if flooding very similar alerts to user. As a business requirement to consider the alerts diversity, our retrieval mechanism combined the diversity of the alerts and ranking score with following two steps:
\begin{itemize}
\item For each metric, detected anomalies are ranked by the ranking score g(p). We take the top anomaly from the list and compare the next highest ranked anomaly with the top one by computing the absolute Pearson correlation coefficient, if the coefficient is less than 0.9, which implies different time series pattern, we include it in the final alerts, if not we compare the next one on the list and so on. Overall, we select two alerts for each metric.
\item Retrieve at most 10 alerts across metrics and rank them with score g(p) from selected anomalies in each metric. 
\end{itemize}

To reduce spam, many business rules are proposed from our user, such as stop alerting duplicated/similar anomalies in recent k days and alerting after k days continuous exceed interval, which achieved good result in eBay.

\section{Validation and Production Performance}
To evaluate our approach, we crowdsourced alert labels from product owners and domain analysts in eBay for a selected set of metrics to collect alert feedbacks (valid alert or not). Also, on our production environment, every alert that we sent out, are triaged and labeled as a valid alert if it is a truly useful and actionable issue with further investigation by the domain analysts.

For the crowdsourced labels, we showed the time series to some colleagues that labeled them independently. Here are the details of our crowdsourced labeling setup: 
\begin{itemize}
\item Each person had the same amount ($15$) of time series to look at out of a pool of $164$ different time-series.
\item Each person was told to label all points that shall trigger alert for each $7$-month (from Jan to July) daily time series.
\item While random, the allocation of time series was designed in such a way that each metric was seen by at least two people.
\end{itemize}

At the end of above alert crowdsourcing collection, we received total $321$ alert feedbacks from $38$ different product owners and domain analysts in eBay.

\subsection{Crowdsourcing Labels Analysis}
\begin{figure}[h]
\centering
\includegraphics[width=0.9\columnwidth]{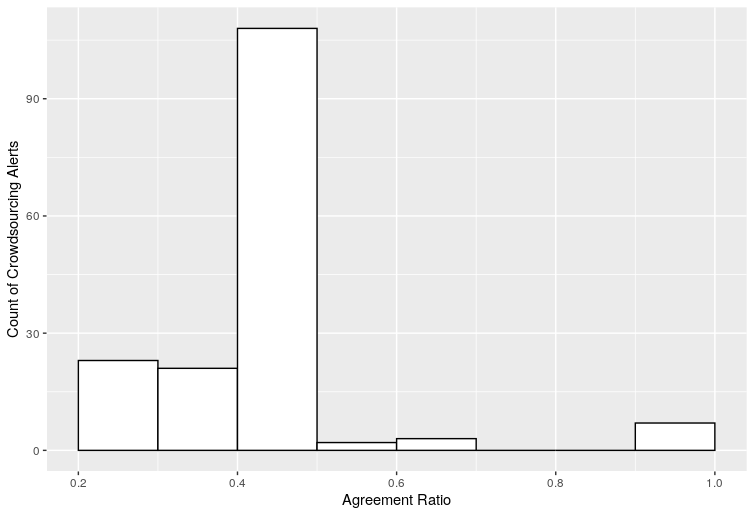} 
\caption{Histogram of Agreement Ratio.}
\label{fig2}
\end{figure}

It turned out that people, when looking at data independently have very differing views about what constituted a valid alert, despite the fact that they were all given the exact same, clear definition of an alert. In Figure \ref{fig2}, in approximately over half of the crowdsourced alert labels there was no common agreement (agreement ratio $\leq 0.50$) as to whether a given data point was an alert or not. Here, the agreement ratio of one alert is defined as the percentage of people who labeled the observation as an alert over the people who saw the time-series.

To understand the lack of agreement situation, we talk with domain analysts to build business rules to reduce spamming.  For example, some domain’s analysts do not expect alerts on just the first day exceed interval since there service exist auto-recovery mechanism, however, they are interested in the $3$ days continuous exceed interval, so we employed the “alerting after k days continuous exceed interval” rule.

\subsection{Validation of Anomaly Detection}
To keep only good quality data, we introduced majority vote rule (agreement ratio $> 0.5$) over the crowdsourced labels, therefore excluding a fair share of data points that were ties. Then, we compared our time series decomposition method with classical method and STL on anomaly detection task using our distribution-free “normal” range. 

In order to weigh in recall more than precision during anomaly detection phase, we choose F2-Score as measurement since it focuses on recalling hundreds of potential alert candidates out of millions. We tuned model parameters with grid search that maximize the F2-Score for each anomaly detection method using different time series decomposer with data during Jan-May 2018, and ran the tuned model over an unseen testing set (Jun-Jul 2018), which we had held aside.

\begin{table}[h]
\centering
\scalebox{0.9}{
\begin{tabular}{lllllllllll}
\toprule
Method & Recall & Precision & F2-Score & Processing Time \\
\midrule
Classical & 64.86\% & 28.92\% & 0.519 & 252.1ms \\
STL & 59.46\% & 37.93\% & 0.534 & 156.8ms \\
Ours & 72.97\% & 38.03\% & 0.616 & 90.3ms \\
\bottomrule
\end{tabular}
}
\caption{Performance comparison on crowdsourcing dataset.}\label{table1}
\end{table}


Besides, we compared speed by using average processing time for every 100 time-series (no parallel computing enabled). The implementation of classical method and STL are the $decompose$ and $stl$ functions in R Package $stats$. In Table \ref{table1}, our decomposer shows better F2-Score performance over eBay Metrics with much faster processing speed.

\subsection{Production Performance}
\begin{figure}[h]
\centering
\includegraphics[width=0.9\columnwidth]{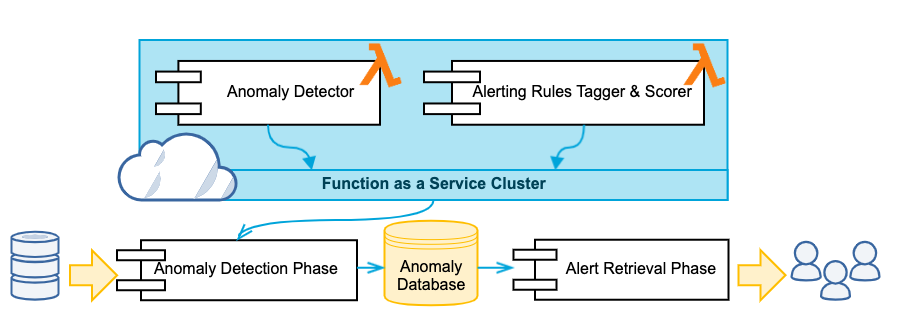}
\caption{Architecture with Function as a Service Cluster.}
\label{fig3}
\end{figure}

Currently, our alert system monitors several million time-series and process hundreds of gigabytes of data every day. we implemented the two-phase solution on our private cloud as several micro-services, such as the data-driven anomaly detector, alerting rules tagger and ranking scorer. In figure \ref{fig3}, Function as a Service (FaaS) framework is adopted to scale up these micro-services as $\lambda$ functions in the anomaly detection phase. Then, the alert retrieval phase leveraged the tags and scores generated in the previous phase to filter and rank anomalies according to domain analysts' settings.

With the 15 servers (each has 8 VCPUs and 16GB RAM) FaaS cluster, Our anomaly detector can handle at least 6.4 thousand time-series per minutes, and we can easily scale up the system by adding additional hardware if needed.

\begin{table}[h]
\centering
\scalebox{0.9}{
\begin{tabular}{lllllllllll}
\toprule
Approach & Weeks & \#Alert & \#Valid Alert & Precision \\
\midrule
First Phase Only & 10 & 164 & 67 & 40.85\% \\
Two Phases & 9 & 118 & 106 & 89.83\% \\
\bottomrule
\end{tabular}
}
\caption{Production performance on valid alert precision.}\label{table2}
\end{table}


To evaluate our alert result, we built an evaluation system to collect alert feedbacks that we sent out. Here, we compared the anomaly detection phase only with the two-phase approach. As in Table \ref{table2}, the extra alert retrieval phase dramatically improved the precision of valid alert.

\section{Conclusion}
To automate product health moving metric alert process at eBay, we developed a two-phase approach to identify anomalies and retrieve valid alerts for different domain users. At anomaly detection phase, we developed the Moving Metric Detector (MMD), which contains a fast and robust time-series decomposition algorithm with better F2-Score performance compared with classical method and STL on anomaly detection task. We leverage Chebyshev’s Inequality to determine a distribution-free “normal” range. To avoid alert spamming and improve alert diversity, we designed the alert retrieval phase with a point-wise ranking model and business rules. Our alerting system is adopted by users across eBay and proved to be an effective way to early alarm issues for business needs.

\section{Acknowledgments}
We thank Rong Song, Christine Wu, Nathan Ni, Zhixuan Jia for their insights and expertise that greatly assisted the research, and continued support by eBay infrastructure organization. We would also like to show our gratitude to Giorgio Ballardin, Woody Zhou, Liren Sun, Jianpeng Xu and Jiahui Ruan for their stimulating discussions. 

\bibliography{references}
\bibliographystyle{aaai}

\end{document}